\documentclass[10pt,twocolumn,letterpaper]{article}

\usepackage[pagenumbers]{cvpr} % To force page numbers, e.g. for an arXiv version

\usepackage{graphicx}
\usepackage{amsmath}
\usepackage{amssymb}
\usepackage{booktabs}

\newcommand{\cvmred}[1]{\textcolor{black}{#1}}
\usepackage[pagebackref,breaklinks,colorlinks]{hyperref}

\usepackage[capitalize]{cleveref}
\crefname{section}{Sec.}{Secs.}
\Crefname{section}{Section}{Sections}
\Crefname{table}{Table}{Tables}
\crefname{table}{Tab.}{Tabs.}

\def\cvprPaperID{*****} % *** Enter the CVPR Paper ID here
\def\confName{CVPR}
\def\confYear{2022}

\begin{document}

%%%%%%%%% TITLE - PLEASE UPDATE
\title{DualSmoke: Sketch-Based Smoke \cvmred{Illustration} Design with Two-Stage Generative Model}

\author{Haoran Xie\thanks{Japan Advanced Institute of Science and Technology, 1-1 Nomi, Ishikawa, 9231292, Japan. E-mail: xie@jaist.ac.jp}\\
JAIST
% For a paper whose authors are all at the same institution,
% omit the following lines up until the closing ``}''.
% Additional authors and addresses can be added with ``\and'',
% just like the second author.
% To save space, use either the email address or home page, not both
\and
Keisuke Arihara\\
JAIST
\and
Syuhei Sato\\
Hosei University
\and
Kazunori Miyata\\
JAIST
}
\maketitle

%%%%%%%%% ABSTRACT
\begin{abstract}
   \cvmred{The dynamic effects of smoke are impressive in illustration design, but it is a troublesome and challenging issue for common users to design the smoke effect without domain knowledge of fluid simulations.} In this work, we propose \cvmred{DualSmoke}, two stage \cvmred{global-to-local} generation framework for the interactive smoke \cvmred{illustration design. For the global stage,} the proposed approach utilizes fluid patterns \cvmred{to generate Lagrangian coherent structure from the user's hand-drawn sketches. For the local stage, the detailed flow patterns are obtained from the generated coherent structure. Finally, we apply the guiding force field to the smoke simulator to design the desired smoke illustration. To construct the training dataset, DualSmoke} generates flow patterns using the finite-time Lyapunov exponents of the velocity fields. The synthetic sketch data is generated from the flow patterns by skeleton extraction. From our user study, it is verified that the proposed design interface can provide \cvmred{various smoke illustration designs} with good user usability. Our code is available at: \href{https://github.com/shasph/DualSmoke}{https://github.com/shasph/DualSmoke}.
\end{abstract}

%%%%%%%%% BODY TEXT
\section{Introduction}

\cvmred{Nowadays, due to the rapid prevalence of social media and personal computing devices, it becomes an emerging and growing topic to help amateur users achieve high-quality content creation. The dynamic effects of 2D illustration play an important role to increase the attraction of various media such as e-books, education materials, and art design. Especially, the design of complex flow design is difficult for common users without the specific domain knowledge and design skills, which are required in commercial design tools, such as PhotoShop and Blender. Previous works adopted kinetic textures~\cite{Kazi2014DracoBL} and elemental dynamics~\cite{Xing2016} to author the dynamical effects, and sketch-based approaches to design fluid motions~\cite{bo2011,hu2019sketch2vf}. However, these approaches may fail to provide the turbulent details in flow motions, especially for smoke design. To solve this issue, we aim to provide an efficient design tool for smoke illustration design.}

\begin{figure}[t]
    \centering
    \includegraphics[width=\linewidth]{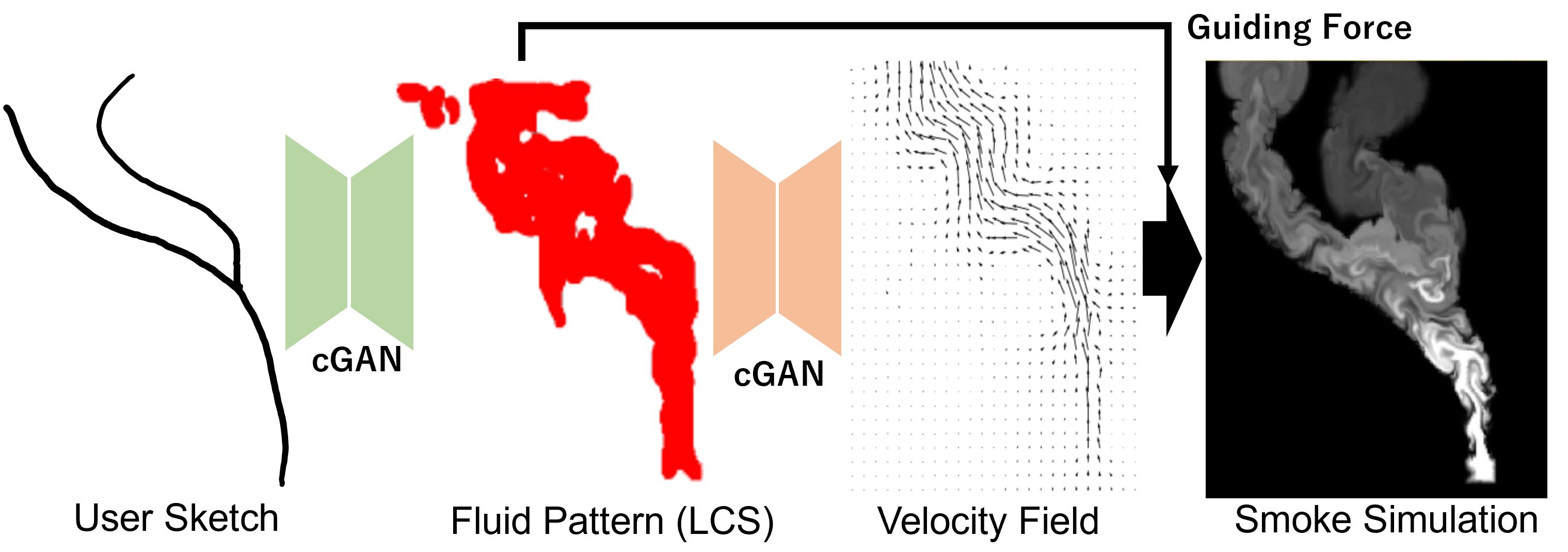}
    \caption{\cvmred{The proposed DualSmoke design framework utilizes fluid patterns to generate velocity field in the generative model.}}
    \label{fig:concept}
\end{figure}

The conventional approaches for fluid control mostly focused on the shape constraints \cite{rasmussen2004directable}. However, a time-consuming process and tedious parameter adjustments are required by users with expert knowledge due to the nonlinear flows. For the keyframe-based approach~\cite{treuille2003keyframe}, the optimization process was designed to explore the appropriate external force, which was normally computationally heavy. In addition, it is not practical to adjust parameters and boundary conditions interactively for smoke simulations. \cvmred{To efficiently design global motions of the fluid, low-resolution simulations were used as input for controlling high-resolution simulation~\cite{Nielsen2010,sato22}.
Low-resolution simulations can be executed at low costs, so users can easily repeat the simulation multiple times to design the motions.
However, designing operations of fluid motions are not still intuitive, because users need to adjust simulation parameters.
For more intuitive design, paths or sketches were introduced as inputs to generate the fluid flows~\cite{Kim2006,kim22sketch2density}.
We follow this idea and our smoke design system is based on sketching operations.
}  

\cvmred{In this work, we propose DualSmoke, a two-stage generation framework with conditional generative adversarial networks (cGAN) for smoke illustration design to meet the users' design intention}, as shown in Figure~\ref{fig:concept}. \cvmred{For the phase of data training,} the proposed framework utilizes Lagrangian coherent structure (LCS) for a given velocity field to represent the main flow in \cvmred{velocity} fields, which is obtained as the ridge of the finite-time Lyapunov exponent (FTLE) field. We then extract the synthetic sketch from the LCS using a heat equation. \cvmred{For the inference phase, we can generate the LCS domain and the velocity field from the input of the user's freehand sketch.} Then, we calculate the guiding force from \cvmred{the generated LCS} and the velocity field \cvmred{and add the force into the smoke simulations.} \cvmred{Note that we set the flow direction as upward in the simulations as common design scenarios.} In order to evaluate the proposed system, we conduct an evaluation experiment to verify the system usability and subjective assessment. In the evaluation experiment, the participants were asked to perform tasks using the proposed system with and without design constraints.

The main contributions of this work are as follows.
\begin{itemize}
    \item We propose \cvmred{DualSmoke}, a two-stage generative model for smoke design considering the flow patterns as structure information in the learning process. 
    \item We constructed the paired training dataset with a pattern-guided smoke simulation where the FTLE and LCS fields are calculated to obtain the synthetic sketches.
    \item We develop a user-friendly design interface with the proposed system for  various design scenarios and verify the proposed user interface with subjective assessment and user experience. 
\end{itemize}

\section{Related Work}

\subsection{\cvmred{Sketch-based Content Creation}}

\cvmred{
Freehand sketching is a naive and efficient way to describe objects and motions for content creation in computer graphics and user interface~\cite{bhattacharjee2020survey}. For geometry modelling, sketch-based interfaces have been proposed for freeform modeling~\cite{igarashi2006teddy}, terrain authoring~\cite{terrian2017}, and character modeling~\cite{sketch2pose}. For physical simulations, sketch input has been adopted in crystal~\cite{Ren2018} and character animations~\cite{sketchcharcter15}. For more intuitive design, paths or sketches were introduced as inputs to generate the fluid flows~\cite{Kim2006,Wu2013}. In this work, we follow this sketch-based idea and adopt the deep learning based approach to guide the smoke design with the rough sketch input to achieve high-quality flow details. }

\cvmred{
Recently, the deep learning based approaches have been explored largely in sketch-based content creation.} Hu et al.~\cite{hu2019sketch2vf} introduced an interactive user interface for fluid design and generated fluid simulations based on the velocity field obtained by cGAN. A similar approach has been used in terrain modeling~\cite{terrian2017}. Yan et al.~\cite{yan2020interactive} used a 3D sketch in an immersive virtual reality environment to generate a splash representation of a liquid with cGAN. \cvmred{For these previous works, it is difficult to achieve high-quality design results. A recent work adopted volumetric density fields in 3D smoke reconstruction using CNN networks~\cite{kim22sketch2density}. However, the sketch input may require fine details of viewpoint and density information, rather than rough hand-drawn sketches. To address these issues, we adopt the global-to-local two-stage deep generative modeling for smoke design. A similar strategy has been utilized in facial image synthesis~\cite{huang2022dualface} and character motion planning~\cite{peng2021two}. This work is the first trial to apply the strategy in smoke design with pattern-guided flow simulations. With the pre-trained generative models, the proposed deep learning based framework can achieve an interactive system without the expensive computation on the fluid simulations.
}

\begin{figure*}[t]
    \centering
    \includegraphics[width=1.0\linewidth]{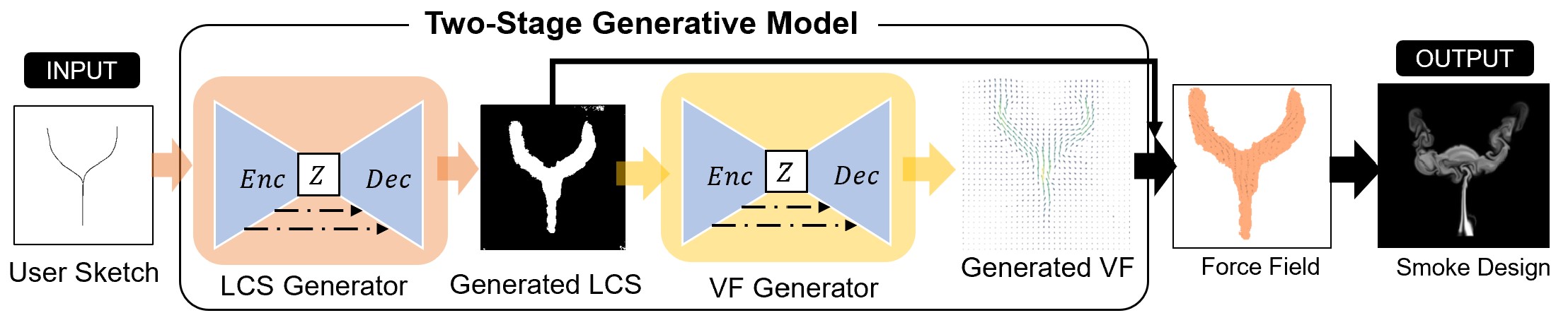}
    \caption{\cvmred{The system framework of the proposed two-stage generative model.}}
    \label{fig:system-abst}
\end{figure*}

\subsection{\cvmred{Control of Fluid and Smoke Simulations}}
For fluid simulation, a stable fluid solver~\cite{stam1999stable} was proposed to be unconditionally stable and has been widely used in research and product communities. Since then, the solver has been improved by other researchers ~\cite{fedkiw2001visual,selle2005vortex}. However, the user needs to repeatedly adjust the parameters and run the simulation to generate the desired flow.

\cvmred{To address this problem, many fluid control methods have been proposed.}
Treuille et al.~\cite{treuille2003keyframe} used the quasi-Newton method to compute the optimal external forces to achieve user-specified keyframes. Rasmussen et al.\cite{rasmussen2004directable} proposed a method to control fluid velocity through control particles. 
\cvmred{Dobashi et al.~\cite{Dobashi2008} controlled the cloud simulation to fit the cloud shape into an input curve. Simulation parameters are automatically adjusted according to the difference between the curve and the cloud shape.
Wu et al.~\cite{Wu2013} introduced the pattern image and computed the driving force based on the image for locally controlling the fluid shape.} 
Tang et al.~\cite{tang2021honey} addressed the lack of parameter space constraints and high dimensionality in the optimization of control forces on high-resolution simulations.
\cvmred{In these methods, users can specify the target shape of the fluid via input curves, images, or 3D meshes, but the flow motion cannot be designed intuitively.}

\cvmred{To efficiently design the fluid flow, some researchers used velocity fields obtained from a low-resolution fluid simulation as an input.
Nielsen et al.~\cite{Nielsen2010} proposed an optimization method to guide the high-resolution simulation according to the low-resolution input.}
Yuan et al.~\cite{yuan2011pattern} proposed a method to control the flow of a high-resolution simulation based on the low-resolution flow by extracting the LCS from the low-resolution flow.
\cvmred{Huang et al.~\cite{Huang2011} also proposed a control method using the low-resolution simulation as a preview. This method can control any properties including velocity fields as well as densities, temperatures, and so on.
Sato et al.~\cite{sato22} formulated the optimization problem in stream function space to reduce the problem size of Nielsen's method~\cite{Nielsen2010}.}
\cvmred{Our system also uses a velocity field and LCS to control the smoke simulation~\cite{yuan2011pattern}, but the velocity field and LCS are obtained via user sketches.}

\subsection{\cvmred{Deep Learning based Fluid Simulation}}
\cvmred{Deep learning based approaches can efficiently create fluid flows, and several methods have recently been proposed.}
Ladicky et al.~\cite{ladicky2015data} achieved a speedup of one order of magnitude compared to position-based simulations at the time by estimating the acceleration of each frame.
\cvmred{Chu et al.~\cite{Chu2017} learned relations between low- and high-resolution velocity fields using CNN (Convolutional Neural Network) for enhancing details of the smoke flow.}
Xie et al.~\cite{xie2018tempogan} used cGAN to generate high-resolution frames from a single frame of a coarse smoke simulation. 
\cvmred{Chu et al.~\cite{Chu2021} proposed a cGAN based model to obtain velocity fields from a single frame of a density field.}
In this study, two cGAN models are trained to generate the LCS with the input of the sketch and the velocity field with the input of the LCS.

\section{System Overview}

The objective of \cvmred{DualSmoke} is to design the velocity field of smoke simulation from the input of user sketches, and obtain a designed smoke flow as system output. To achieve this goal, the proposed system framework of the interactive smoke design \cvmred{adopts a two-stage generative model with the generated LCS and VF fields} as shown in Figure \ref{fig:system-abst}. 

For the two-stage generative model, two generative model networks are adopted to learn the relationships between the LCS and the synthetic sketch data denoted as LCS Generator in Figure \ref{fig:system-abst}, and between the velocity field and the LCS denoted as \cvmred{Velocity} Field (VF) Generator. \cvmred{After generating the velocity field and the LCS flow patterns from this model, }we calculate the guiding force field from the generated LCS field and the velocity field from the input of the user sketch. By using the conventional fluid simulator, the smoke simulation can be generated with the guiding force fields in real time. Finally, we develop the design interface for the smoke simulation using the proposed framework. \cvmred{For the design interface, we adopt both smoke control lines and obstacle lines in the smoke design: the control lines represent the user sketch in the fluid domain for the proposed generative models; the obstacle lines represent the rigid objects which are utilized in boundary conditions of the smoke simulation. }

\section{Training Preprocess}

\cvmred{For the training preprocess, we collect the training dataset for the deep learning-based generative models. From a smoke simulation with the given velocity fields, we calculate the FTLE structure and then obtain the LCS fields using a Gaussian mixture model (GMM). Based on the obtained LCS fields, we calculate the synthetic sketches using the skeleton extraction method with a heat equation. }

\subsection{Smoke Simulation}

\cvmred{For a smoke simulation at any timestep $t$, }the materials are advected by the velocity field $u$ and the pressure field $p$. Assuming that $u$ and $p$ at the initial time $t = 0$ are known, the variation of these quantities can be solved using Navier–Stokes equations, as follows:

\begin{equation}
    \nabla\cdot{\bf u}=0
    \label{eq:continue}
\end{equation}
\begin{equation}
    \frac{\partial{\bf u}}{\partial t}=-({\bf u}\cdot\nabla){\bf u}-\frac{1}{\rho}\nabla p+\nu\nabla^2{\bf u}+{\bf f}
    \label{eq:navier}
\end{equation}
\begin{equation}
    {\bf u} \cdot {\bf n} = 0
    \label{eq:boundary}
\end{equation}

\noindent
where $\nu$ is the kinematic viscosity, $\rho$ is the flow density, and $\bf f$ is the external force. The solid wall boundary condition is adopted for the smoke flow around the obstacle domain in Equation~\ref{eq:boundary}, where $\bf n$ denotes the normal vector on the solid boundary. \cvmred{In this work, we used the semi-Lagrangian method \cite{stam1999stable} for advection term, and the MacCormack method \cite{selle2008unconditionally} to suppress the numerical diffusion on grids.}

\begin{figure}[t]
\centering
\includegraphics[width=\linewidth]{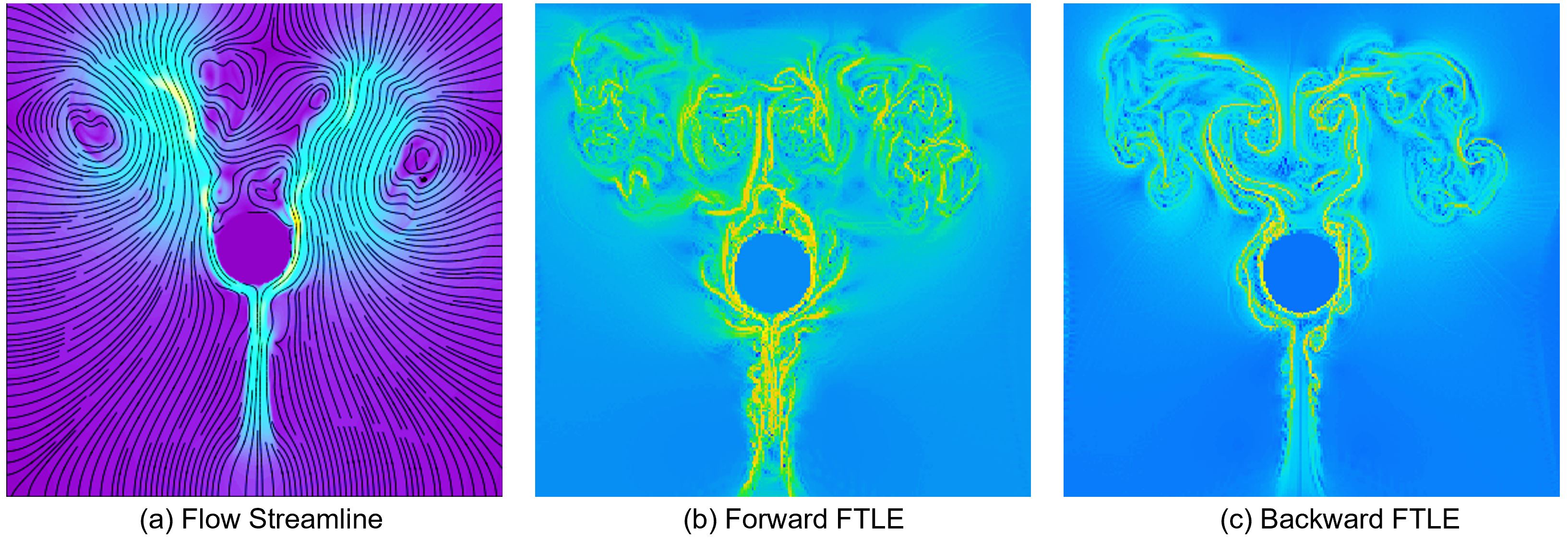}
\caption{Flow patterns with forward FTLE (b) and backward FTLE (c) with a \cvmred{velocity} field (a).}
\label{fig:ftle-diff}
\end{figure}

\subsection{Finite-Time Lyapunov Exponent}
The FTLE is calculated by tracing a particle placed at the computation point in the velocity field\cvmred{. We obtain the FTLE within the integration interval based on the change in the particle position at the end~\cite{shadden2005definition}.} 
The trajectory of a particle advanced from its initial position $t_0$ by an integration time $T$ is defined by the following flow map $\Phi$.
\begin{equation}
    \Phi_{t_0}^{t_0+T}({\bf p}) = {\bf p}(t_0+T)
    \label{eq:flowmap}
\end{equation}
where ${\bf p}(t)$ is the position of the particle at time $t$ and $\bf u$ is the velocity. For the flow map, the local deformations are computed using the Cauchy–Green deformation tensor as follows:
\begin{equation}
\label{eq:cauchy-green}
    %\Delta\coloneqq{}(\nabla_{t_0}^{t_0+T}({\bf p}))^T(\nabla_{t_0}^{t_0+T}({\bf p}))
    \Delta = \mathbf{M}^*\mathbf{M}, \mathbf{M}=\frac{d\Phi_{t_0}^{t_0+T}({\bf p})}{d{\bf x}}
\end{equation}
\cvmred{where $\mathbf{M}^*$ is the transpose of $\mathbf{M}$ and $\Delta$ is a $2\times2$ matrix for a 2D flow in this work. The FTLE value is defined as a time-dependent scalar quantity using the largest eigenvalue of the matrix, $\lambda_{max}$ as follows. }
\begin{equation}
\label{eq:ftle}
    \sigma_{t_0}^T({\bf p})=\frac{1}{|T|}\ln{\sqrt{\lambda_{max}(\Delta)}}
\end{equation}
\cvmred{where $\ln$ is the natural logarithm. Note that the FTLE fields are different according to the Lagrangian particles' movements. When $T>0$, the particle moves forward and the flow is repulsive (Figure \ref{fig:ftle-diff}(b)); when $T<0$, the  particle moves backward and the flow converges (Figure \ref{fig:ftle-diff}(c)). In this work, we use backward FTLE to find the convergence point of the flow and extract the flow patterns. }

\subsection{\cvmred{GMM-based Lagrangian Coherent Structure}}
\begin{figure}[tb]
    \centering
    \includegraphics[width=\linewidth]{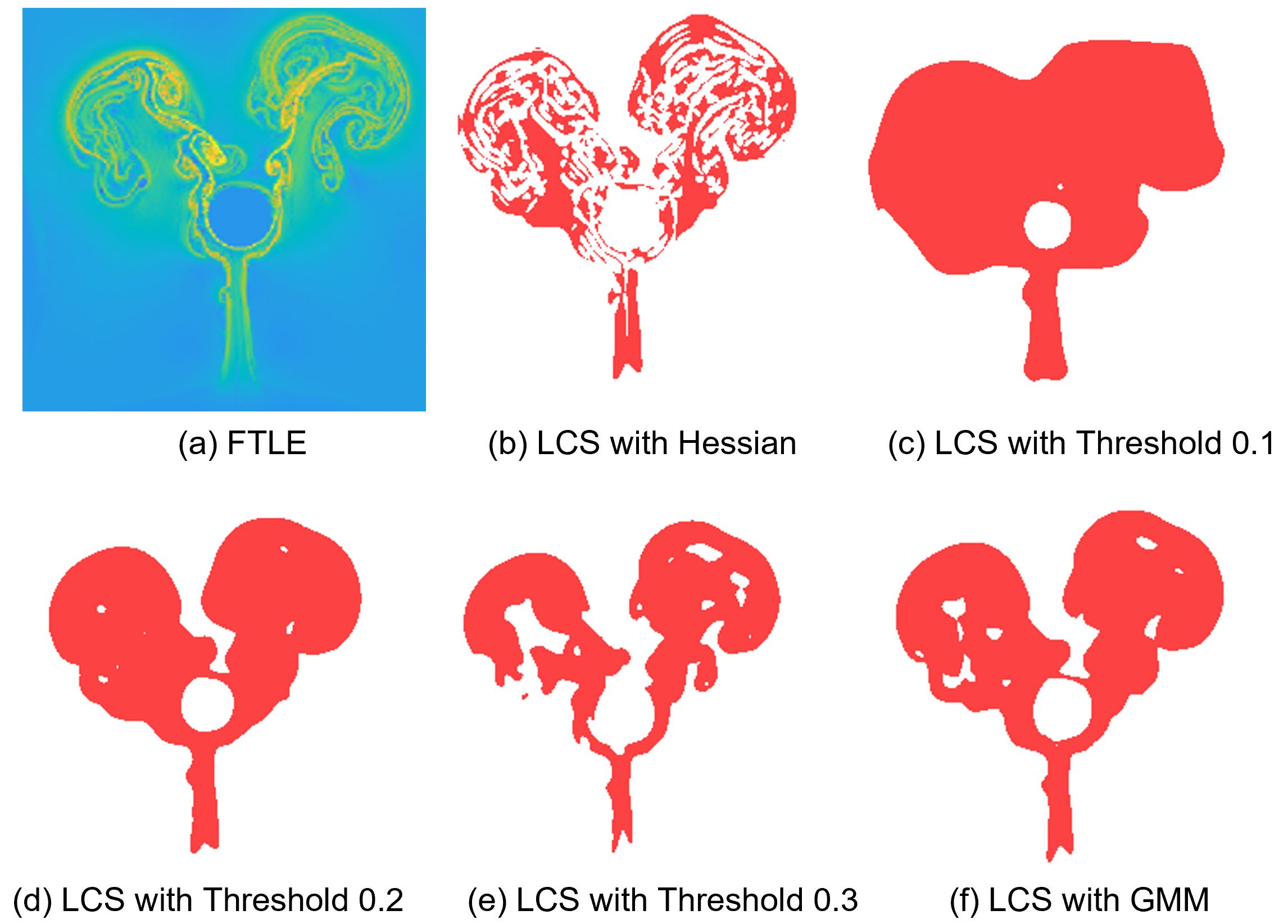}
\caption{Comparison results of the LCS abstract approaches: (a) Input of FTLE; (b) Hessian; (c)-(e) Threshold \cite{yuan2011pattern}; (f) GMM (ours). }
\label{fig:lcs-diff}
\end{figure}

\cvmred{The LCS reveals the flow pattern is obtained by extracting the ridge of the maxima of the calculated FTLE field. Figure~\ref{fig:lcs-diff} shows the LCS extracted image using a Hessian matrix and threshold-based approaches \cite{ferstl2010interactive,yuan2011pattern}. A certain threshold value may achieve good extraction results from FTLE fields. However, the extraction with the same threshold value caused artifacts in dataset construction because the optimal threshold values were varied for different FTLE fields. }Therefore, we propose a GMM to obtain the optimal threshold value for FTLE fields in this study.

\cvmred{A GMM is used to extract the LCS from the computed FTLE field \cvmred{which classifies the given data into multiple normal distributions}, as shown in Figure~\ref{fig:lcs-diff}(f).  In this work, the FTLE field is divided into two classifiers, and the larger mean values of each class are used as the threshold values. We applied a Gaussian filter before clustering due to the noise generated in the integration calculation of FTLE fields.}
\begin{figure}[t]
    \centering
    \includegraphics[width=1.0\linewidth]{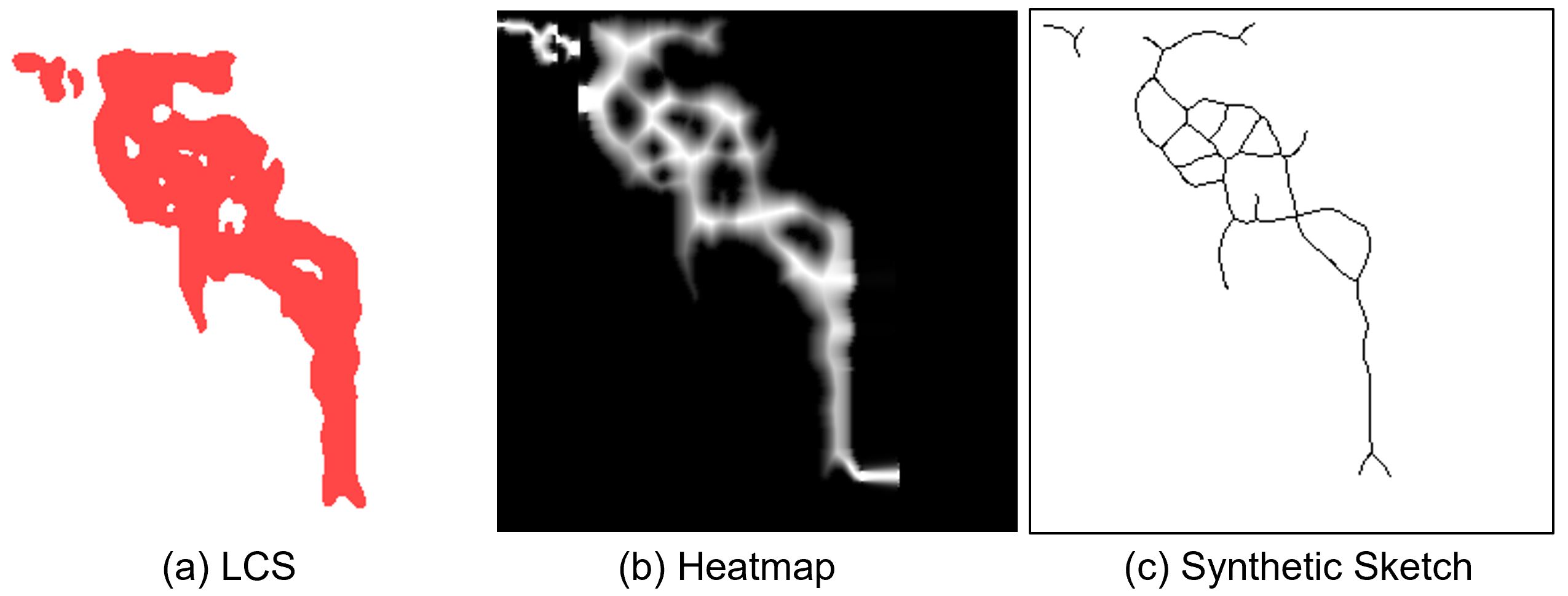}
    \caption{Synthetic sketch generation using the proposed skeleton extraction.}
    \label{fig:sketch}
\end{figure}

\begin{figure*}[t]
    \centering
    \includegraphics[width=0.8\linewidth]{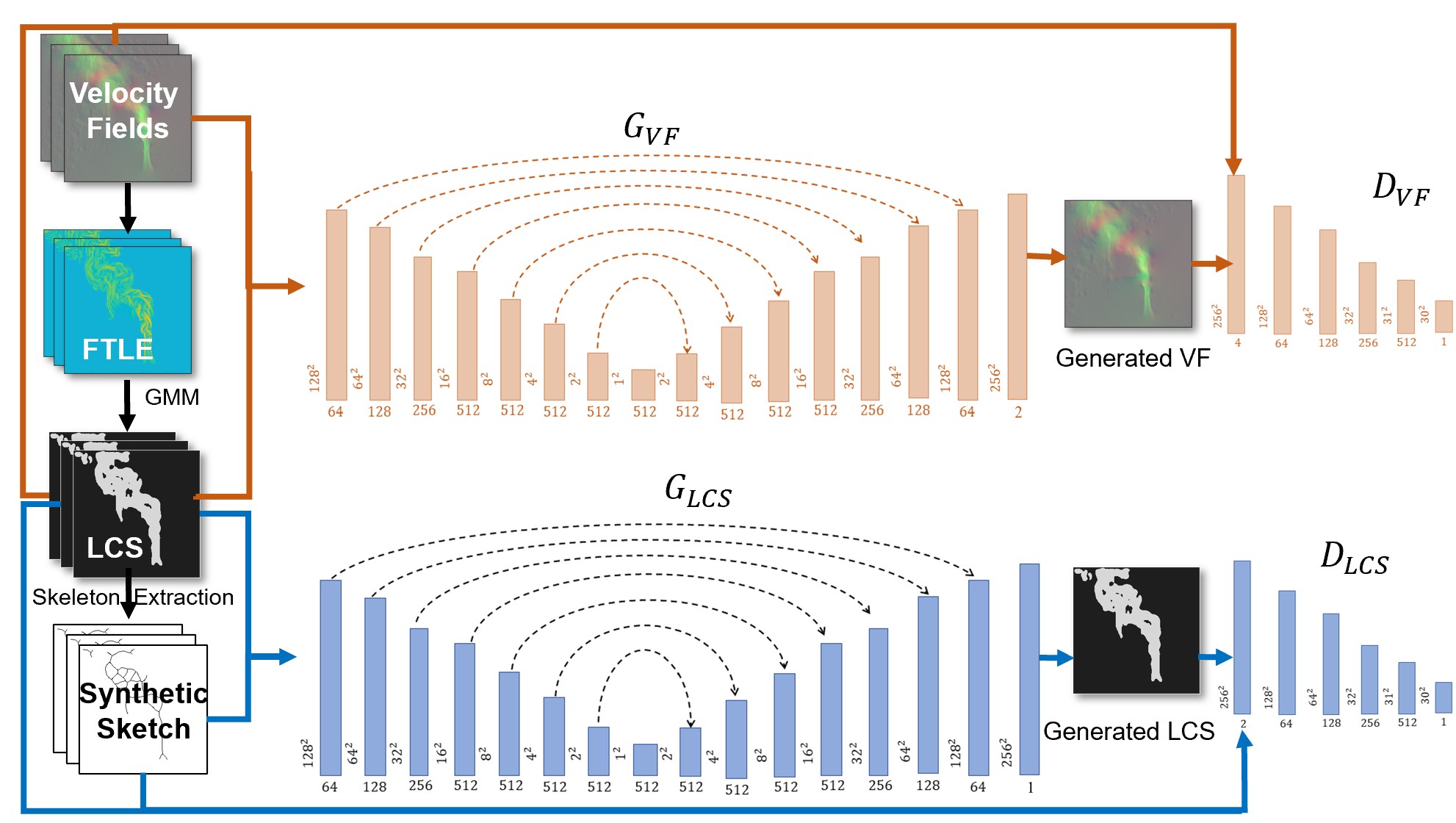}
    \caption{\cvmred{Network structure of our proposed two-stage generative model. The training data includes velocity field, LCS, and the synthetic sketch on the left hand. There are two generators ($G_{LCS}$ and $G_{VF}$) with U-Net structure and two discriminators ($D_{LCS}$ and $D_{VF}$) with PatchGAN structure. The numbers on the left of blocks represent the spatial size of the network layer and the ones below layer blocks for the number of layers.}}
    \label{fig:network}
\end{figure*}

\subsection{Synthetic Sketch}
Skeleton extraction is proposed to generate a synthetic sketch from the extracted LCS (Figure~\ref{fig:sketch}). The important issue in skeleton extraction is to preserve the topological structure of the input 2D shape and the geometric features of the original shape. To avoid artificial lines and the post-process, such as pruning of the generated branches~\cite{tagliasacchi2012mean,au2008skeleton}, we use the heat equation~\cite{GAO201899} to extract skeletons. We first place a heat source in the contour of the 2D LCS image and block its propagation outside the extraction region. Finally, we can extract the ridge of the temperature map obtained by proceeding with the thermal simulation.

\section{Generative Model}
For our two-stage generative model, we first discuss a two-stage strategy with paired training data. The network structures of both generative models are then described.

\subsection{Two-Stage Strategy}
For a conditional generative model with a deep learning approach, it is challenging to reveal the structural information in the training process. Especially for complex domains such as smoke simulations, there are detailed structures in the velocity fields such as turbulent details. To solve this issue, we explicitly adopt a two-stage strategy in the generation of a smoke simulation. The global features of LCS skeletons are generated from initial sketch inputs, and the local details of velocity fields are generated from the LCS fields.    

\begin{figure}[t]
\centering
\includegraphics[width=0.95\linewidth]{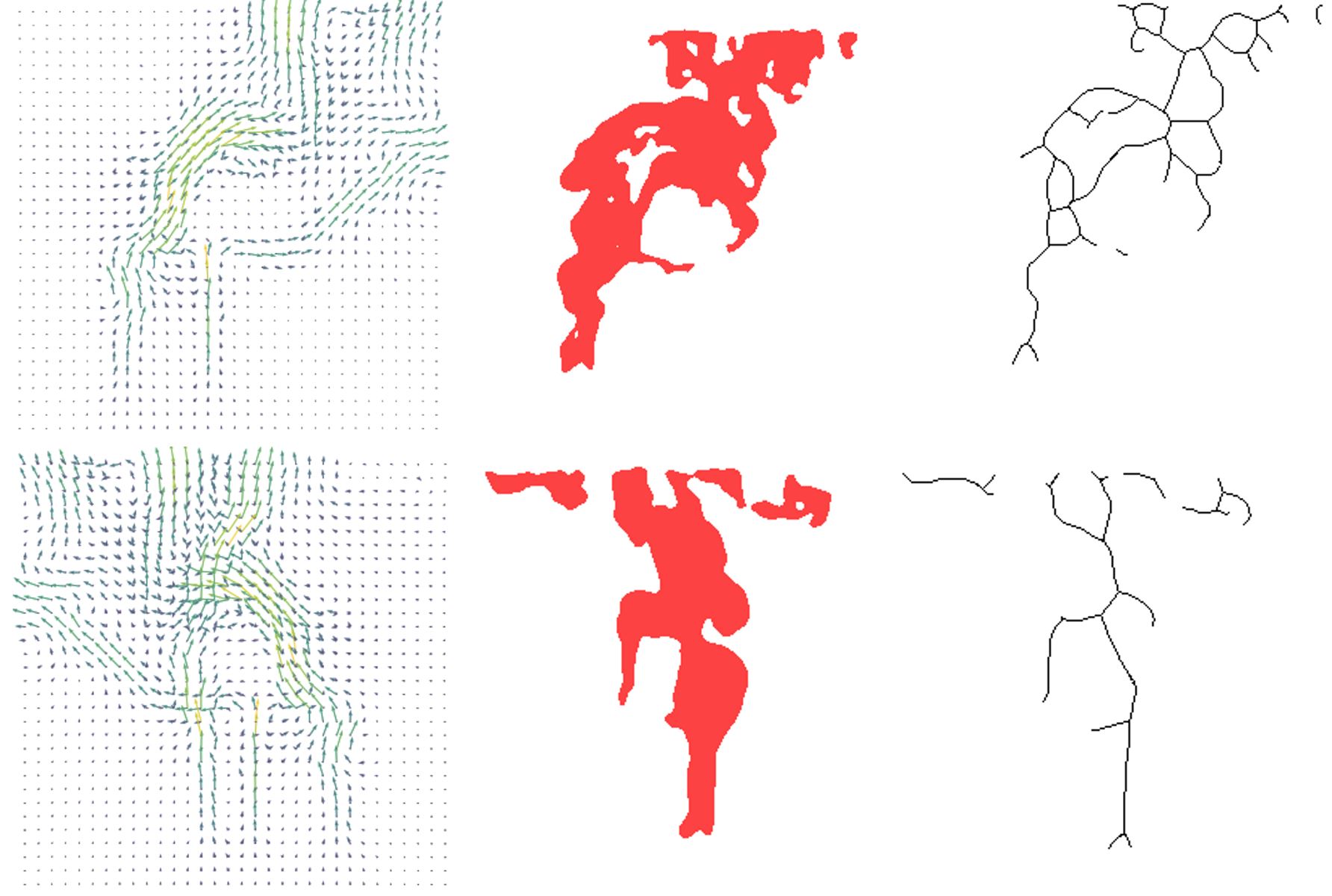}
\caption{Examples of the paired training data used in our work. From left to right: the \cvmred{velocity} field, the LCS, and the synthetic sketch.}
\label{fig:train-data}
\end{figure}

\cvmred{Figure~\ref{fig:network} shows the network structure of the proposed DualSmoke global-to-local strategy for generative models.} In our training dataset, we collect the paired synthetic sketch and the LCS data for the LCS generator, and the LCS and velocity fields for the VF generator. The training data consists of the velocity fields generated by the smoke simulation, the LCS, and the synthetic sketches, as illustrated in Figure~\ref{fig:train-data}. The sketch data in the training data is obtained by extracting the skeleton from the LCS. 

\subsection{Network Structure}
In this study, we use the cGAN-based generative structures~\cite{pix2pix2017} for LCS generator $G_{LCS}$ from hand-drawn sketches $\boldsymbol{S}$, and VF generator $G_{VF}$ from the generated LCS field $\boldsymbol{L}$ (Figure~\ref{fig:system-abst}). The generators are based on the U-Net structure with skip connections between the decoder and encoder networks~\cite{ronneberger2015u}. The discriminators $D_{LCS}$ and $D_{VF}$ are constructed similar to PatchGAN~\cite{pix2pix2017} \cvmred{as shown in Figure~\ref{fig:network}}. The loss function of the LCS generator is designed with cross-entropy, as follows:
\begin{equation}
\begin{split}
\min_{G_{LCS}}\max_{D_{LCS}}
\{ \mathbb{E}_{(\boldsymbol{S}, \boldsymbol{L})}
[log D_{LCS}(\boldsymbol{S}, \boldsymbol{L})] \\
 + \mathbb{E}_{\boldsymbol{S}}
[log (1-D_{LCS} (\boldsymbol{S}, G_{LCS}(\boldsymbol{S})))]\}
\end{split}
\label{LCgan_minimax}
\end{equation}
\cvmred{where $\mathbb{E}_{(\boldsymbol{S}, \boldsymbol{L})}$ and $\mathbb{E}_{\boldsymbol{S}}$ denote the expected values over sketch $\boldsymbol{S}$ and generated LCS $\boldsymbol{L}$ fields, respectively.}
The loss function of the VF generator with the generated velocity field $\boldsymbol{V}$ is given as follows.
\begin{equation}
\begin{split}
\min_{G_{VF}}\max_{D_{VF}}
\{ \mathbb{E}_{(\boldsymbol{L}, \boldsymbol{V})}
[log D_{VF}(\boldsymbol{L}, \boldsymbol{V})] \\
 + \mathbb{E}_{\boldsymbol{L}}
[log (1-D_{VF} (\boldsymbol{L}, G_{VF}(\boldsymbol{L})))]\}
\end{split}
\label{LCgan_minimax}
\end{equation}
\cvmred{where $\mathbb{E}_{(\boldsymbol{L}, \boldsymbol{V})}$ and $\mathbb{E}_{\boldsymbol{L}}$ denote the expected values over LCS $\boldsymbol{L}$ and generated velocity field $\boldsymbol{V}$ fields, respectively.} The deep-learning model was trained with a batch size of 64 and 100 epochs. The optimization algorithm used was Adam for both the generator and the  discriminator, with a learning rate of 0.0002.

\section{Smoke Design}
For the runtime smoke design, we calculate the force fields with the generated results from the aforementioned generative model that are used in the conventional fluid solver for simulation control. In addition, we developed a design interface for the interactive smoke design with the proposed system. 

\subsection{Guiding Force Field}
\label{section:guide-force}
In the smoke simulations, the flow behavior can change significantly with the change in simulation resolution and the addition of turbulence. To solve this issue, we adopted the guided smoke simulation using the velocity and LCS fields generated from the generative models~\cite{yuan2011pattern}. In the Navier–Stokes equations, we represent the external force ${\bf f}$ in Equation~\ref{eq:navier} as follows:

\begin{equation}
    {\bf f}=\alpha{\bf z}+\bf F(\bf p)
    \label{eq:extForce}
\end{equation}
 where the first item is the buoyancy force, ${\bf z}=(0,1)$ and $\alpha$ is a positive constant; $\alpha=0.025$ was used for most cases in this work. The second item $\bf F(\bf p)$ denotes the guiding force, which can be calculated from the generated LCS and velocity field. 

\begin{figure}[tb]
    \centering
    \includegraphics[width=1.0\linewidth]{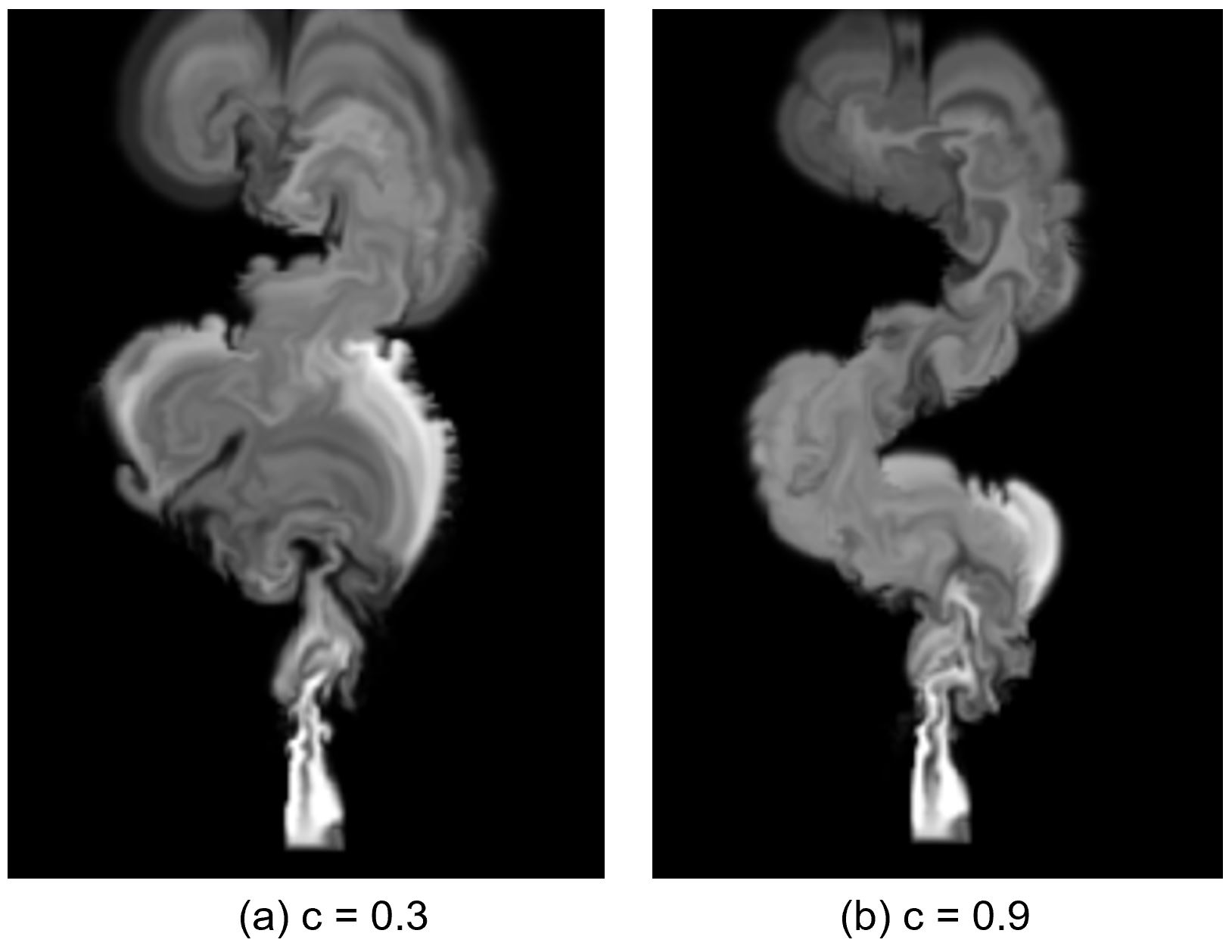}
\caption{Smoke simulation results with different scaling parameters.}
\label{fig:scale-diff}
\end{figure}

\begin{figure}[tb]
    \centering
    \includegraphics[width=1.0\linewidth]{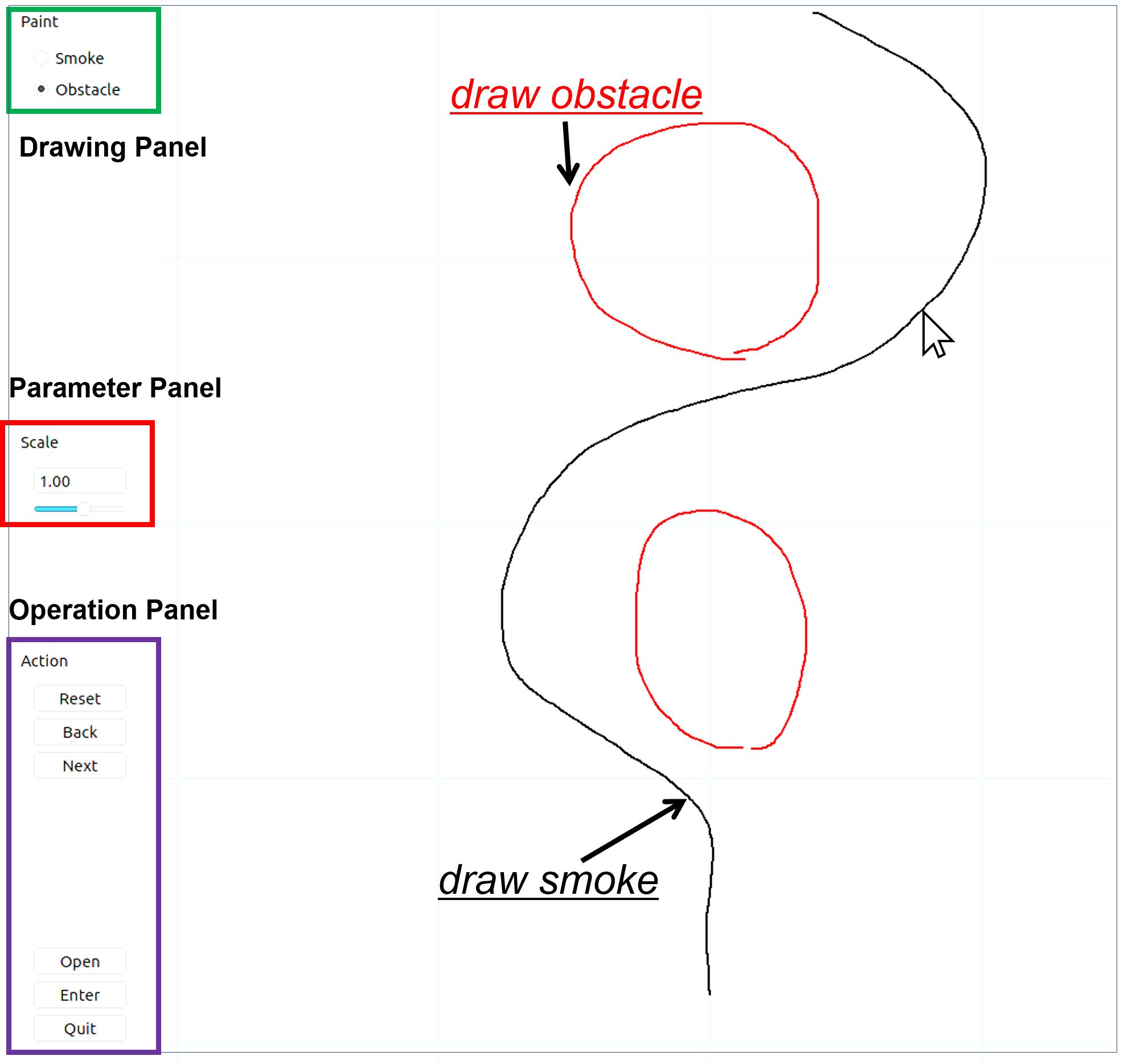}
    \caption{User interface for interactive smoke design.}
    \label{fig:user-interface}
\end{figure}

The region of the velocity field to be guided is determined based on LCS region $\Omega$ as force guidance. The guiding force $\bf F({\bf p})$ at point $\bf p$ is calculated as follows.
\begin{equation}
    \bf F(\bf p)=
    \begin{cases}
        c\frac{1}{\delta t}({\bf u}_{G}-{\bf u}_{S}) & \text{if {\bf p} $\in\Omega$;}\\
        0 & \text{if {\bf p} $\notin\Omega$.}
    \end{cases}
\end{equation}
where ${\bf u}_{G}$ is the generated velocity field and ${\bf u}_{S}$ is the velocity field in the simulation. The constant $c$ is a scaling parameter for the guiding force and indicates the degree to which the simulation follows the guiding force. \cvmred{Due to the trade-off between high-fidelity to freehand sketches and the naturalness of simulation results for sketch-based design, we leave the design freedom to the users to adjust $c$ value (the default value is 1.0)}. When $c$ is small, the simulation is less constrained by the velocity field and is more sensitive to external forces, such as buoyancy. However, as $c$ becomes larger, the influence of external forces becomes smaller, and the motion tends to be sharper. This can be adjusted by the user in the parameter panel of our user interface (Figure~\ref{fig:user-interface}). \cvmred{Because of the usage of the pre-trained model (around 11ms to generate the simulation result), the proposed DualSmoke system was operated in real time so that the users can adjust $c$ value intuitively}. An example of the difference in smoke motions with different scaling parameter $c$ is shown in Figure~\ref{fig:scale-diff}.

\subsection{User Interface}

The design interface can support user drawing and control with interactive simulation feedback, as shown in Figure~\ref{fig:user-interface}. Our interface provides the main canvas and three functional panels. For the drawing panel, a radio button is provided to toggle between smoke control (smoke) and an obstacle (obstacle) line. The control line is black and the obstacle is red. Note that the control line is used as the sketch input of our proposed two-stage generative model, and the solid obstacle is handled in the solid wall boundary condition in Equation~\ref{eq:boundary}.

For the parameter panel, we provide the scaling parameter modification to change the scales of the guiding force. For the operation panel, the operational functions are provided, such as backward, forward, and reset functions. The user can load the previously saved sketches for editing purposes and can start the smoke simulation by clicking the ``Enter" button to confirm the designed results in real time.

\begin{figure*}[t]
    \centering
    \includegraphics[width=\linewidth]{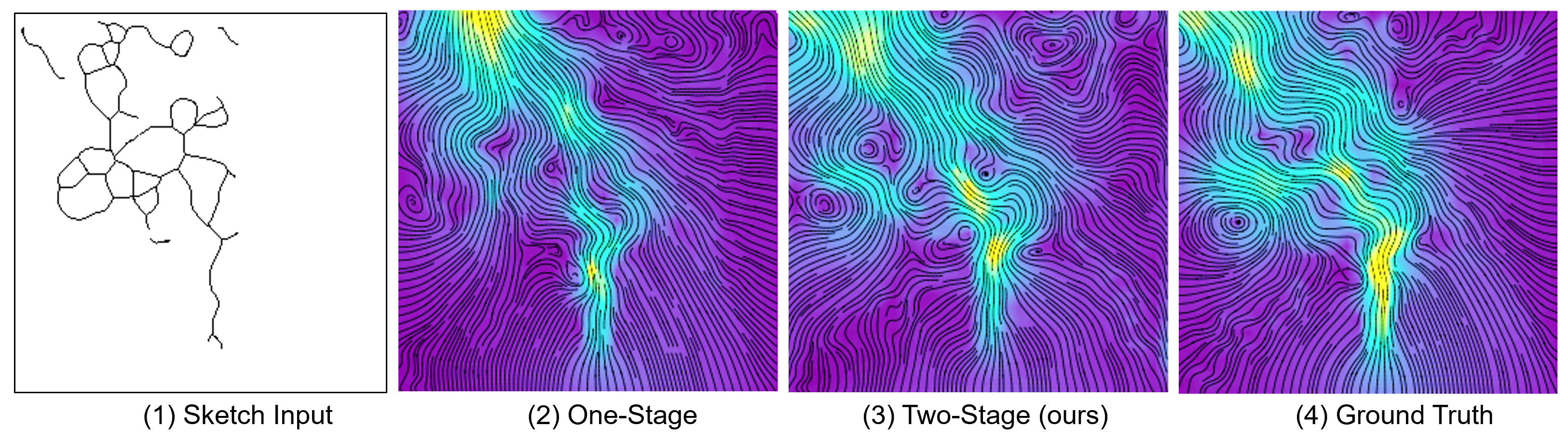}
\caption{Comparison results for the one-stage generative model (2), our proposed framework (3), and the ground truth (4). \cvmred{Note that the sketch input (1) is the synthetic sketch extracted from the ground truth. The lines and colors indicate the streamline and magnitude of the velocity field, respectively. }}
\label{fig:compare}
\end{figure*}

\begin{figure*}[t]
    \centering
    \includegraphics[width=\linewidth]{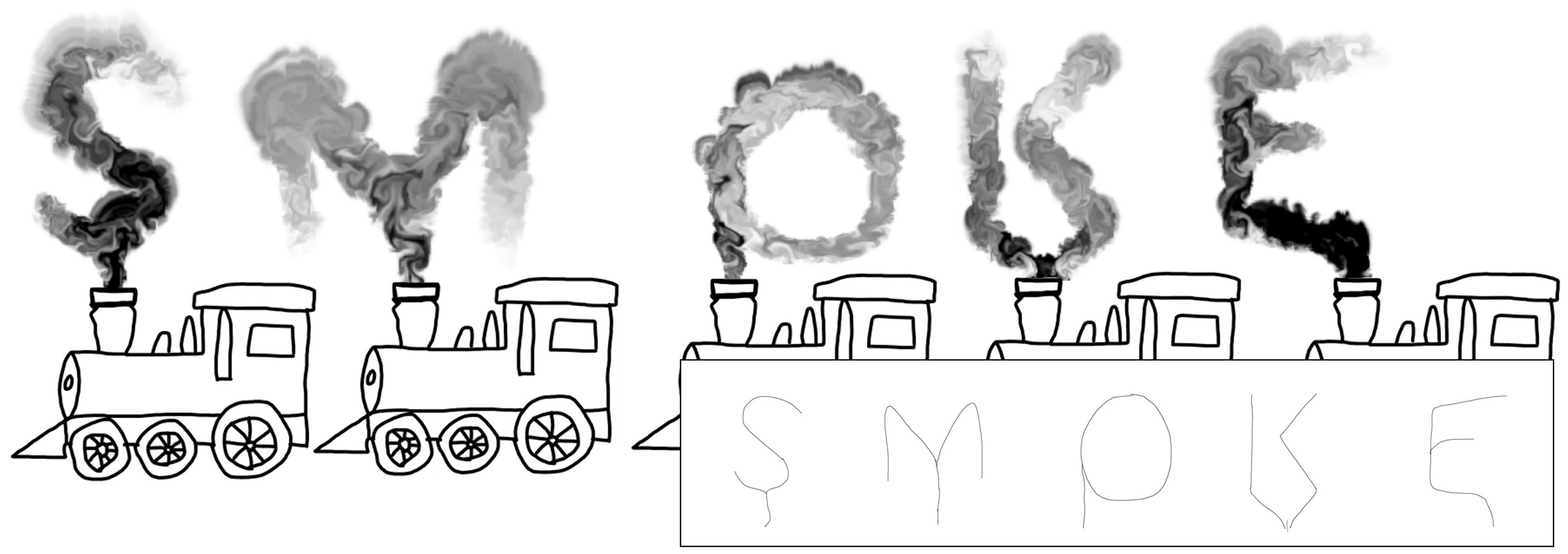}
\caption{Smoke illustration design of steam locomotive in ``SMOKE" shapes. The bottom-right sub-figure is the input of user sketches.}
\label{fig:train}
\end{figure*}

\section{Implementation Details}
The proposed \cvmred{DualSmoke} interface is implemented by PyQt, and the smoke simulation is driven by the mantaflow framework~\cite{mantaflow-url}. The generative models were trained on RTX3090 and Intel Xeon W-2223@3.60GHz. \cvmred{In our system prototype, all smoke simulations were conducted on $256 \times 256$ grids. For the generation of the LCS and the velocity field using the pre-trained model, the generation time from user sketch to LCS was 8 ms, and the generation time from the LCS to the velocity field was about 3 ms. The data learning process spent around 6 hours for LCS generator, and 18 hours for VF generator.}

\subsection{Training Data}
We adopted the grid-based solver for smoke simulations to obtain the velocity fields for training data. The semi-Lagrangian method was used for advection calculation. Simulations for dataset construction were run in an environment where the outflow position was determined by random x-axis and fixed y-axis coordinates, and a square area of a uniform randomly oriented velocity field was placed in the center. The velocity field and the locations of the sources made the flow in each simulation more random, and the data became diversified. The simulation interval is $\Delta t=0.1$, that is 1000 frames per 100 seconds. The velocity field used as training data was that of the 1000-frame point, and the FTLE was calculated as the backward motion from the 1000th frame, with an integration interval of $T=2.5$, or 25 frames. For the standard deviation of the mixed Gaussian model, $\sigma=1.0$ was used. \cvmred{The dataset consisted of 1374 paired training data and 344 paired test data.}

\subsection{LCS Calculation}
To calculate the LCS from the given velocity field, we first compute the FTLE by tracing a Lagrangian particle placed with the velocity field during the integration interval $T$. For the 2D fluid domain in grid point $(i,j)$, the four particles are placed at ${\bf p}_{left}=(i-\tau,j)$, ${\bf p}_{right}=(i+\tau,j)$, ${\bf p}_{down}=(i,j-\tau)$, and ${\bf p}_{up}=(i,j+\tau)$, where $\tau$ is the infinitesimal distance (in this work, we use 0.1 for a unit grid length of 1). The four particles are traced in the integration interval $T$ using the velocity field of each frame. Equation~\ref{eq:cauchy-green} is computed numerically, with the traced particles as ${\bf p}'_{left}$, ${\bf p}'_{right}$, ${\bf p}'_{up}$, and ${\bf p}'_{down}$, respectively.
\begin{equation}
    \frac{d\Phi_{t_0}^{t_0+T}({\bf p})}{d{\bf x}}= \\
    \begin{pmatrix}
        \frac{x({\bf p}'_{right})-x({\bf p}'_{left})}{2\tau} & \frac{x({\bf p}'_{up})-x({\bf p}'_{down})}{2\tau} \\
        \frac{y({\bf p}'_{right})-y({\bf p}'_{left})}{2\tau} & \frac{y({\bf p}'_{up})-y({\bf p}'_{down})}{2\tau} \\
    \end{pmatrix}
\end{equation}
Finally, the FTLE is obtained by calculating Equation~\ref{eq:cauchy-green} from the maximum eigenvalue in Equation~\ref{eq:ftle}. The fourth order Runge–Kutta method was used for tracing, and linear interpolation was used to interpolate the particle positions. The calculated FTLE fields are classified into two normal distributions by a GMM, and LCS is calculated based on the threshold values.

\section{System Results}

\subsection{Ablation Study}

\cvmred{To verify the proposed network structure, we compared the  generative model with and without LCS flow patterns.} Figure~\ref{fig:compare} shows the comparison results among our two-stage approach, the one-stage generative model \cvmred{without LCS. For the network verification, we adopted the extracted skeleton from the ground truth as the sketch input, rather than the user sketches. Therefore, the sketch complexity has no apparent influence on the network structure.} For the one-stage generative model, we trained the model with paired synthetic sketches and \cvmred{velocity} fields in cGAN network. We found that the color distribution in our proposed approach is closer to ground truth, which indicates that the global flow features are captured well. For turbulent details, we calculated L1 and L2 distances and the cosine similarity of the normalized velocity fields in the one-stage model and our two-stage generative model, as shown in Table 1. Our method has smaller L1 and L2 distances of errors for whole and LCS domains. Regarding the cosine similarity, our approach achieved better performance in the LCS domain as indicated by the global features. Therefore, we verified that the proposed method can achieve better both global fluid patterns and local turbulent details than the one-stage model.

\begin{table}[hbtp]
  \caption{Comparison results for one-stage and two-stage frameworks.}
  \label{table:Error}
  \centering
  \begin{tabular}{lcr}
    \hline
    & One-Stage & Two-Stage (Ours)\\
    \hline \hline
    L1 & 0.073 & 0.071\\
    L2 & 0.014 & 0.011 \\
    L1 (LCS) & 0.037 & 0.031\\
    L2 (LCS) & 0.01 & 0.007\\
    Cosine Similarity & 0.677 & 0.644\\
    Cosine Similarity (LCS) & 0.612 & 0.665\\
    \hline
  \end{tabular}
\end{table}

\subsection{Application Design}

\begin{figure}[tb]
    \centering
    \includegraphics[width=\linewidth]{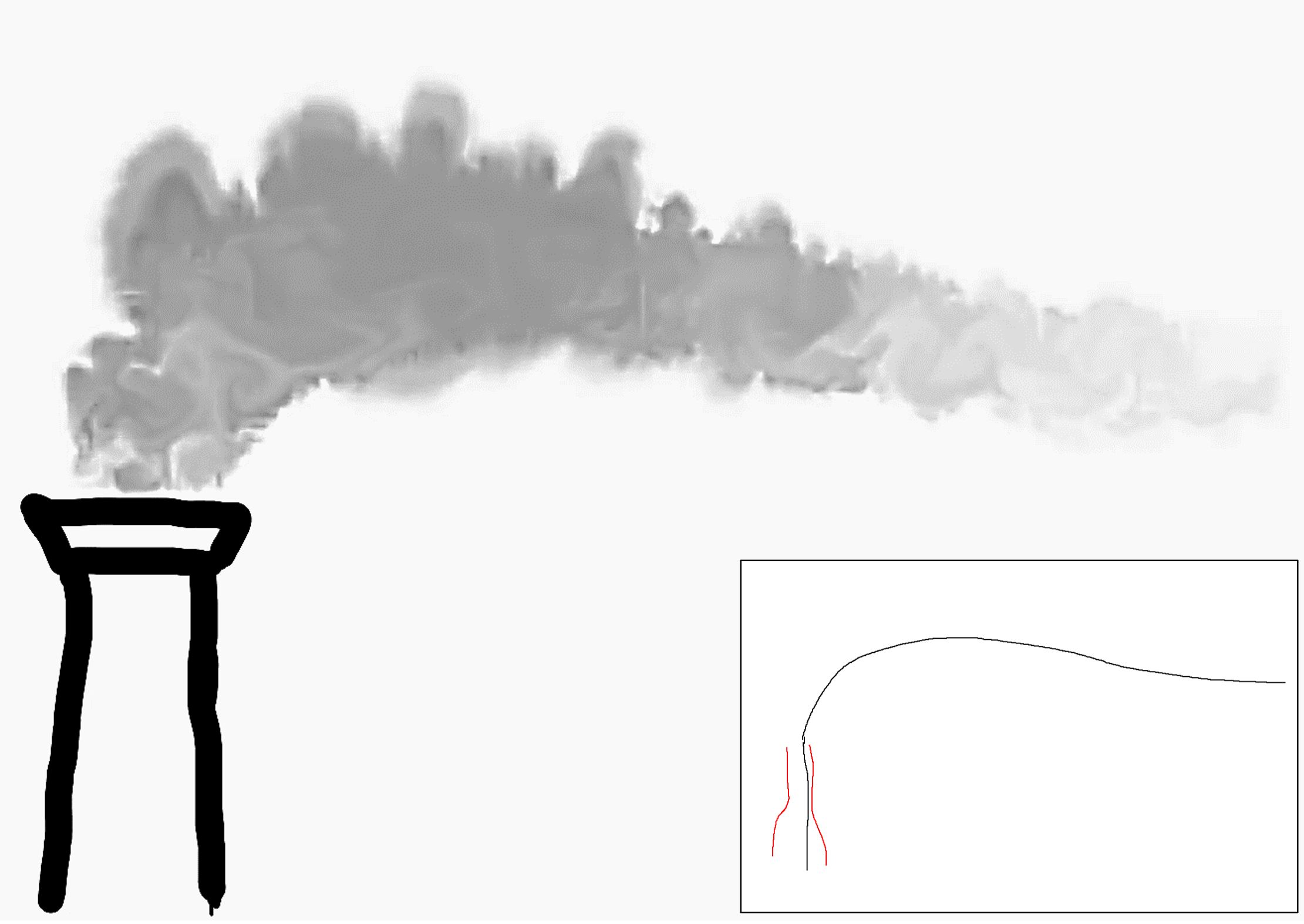}
\caption{Smoke design from a chimney from user sketches. The red lines in the bottom-right figure denote the obstacle as a chimney. The black line was designed for smoke control.}
\label{fig:chimney}
\end{figure}

The proposed \cvmred{DualSmoke} can be applied in various applications and has various practical uses. 
Figure~\ref{fig:train} shows a smoke illustration with the designed ``SMOKE" text shapes. The scaling parameters used are $c=1.3$ for the S-shape, $c=0.8$ for the M, $c=1.8$ for the O, $c=0.85$ for the K, and $c=1.47$ for the E. We can observe that the smoke may accumulate in the curved lines because the smoke can gain momentum due to the buoyancy and guiding force. Figure~\ref{fig:chimney} shows the illustration design in which smoke from a chimney (here, $c=0.8$) . It is difficult for a designer to control complex smoke performance. With the help of the proposed system, users can simply draw the smoke flow with a single stroke. The simulated result is physically correct with turbulent details, which cannot be achieved by previous sketch-based flow design \cite{bo2011,hu2019sketch2vf}. The proposed design framework enables the smoke generation in not only single stroke characters but also a variety of characters by designing the flow patterns.

\begin{figure}[tb]
    \centering
    \includegraphics[width=\linewidth]{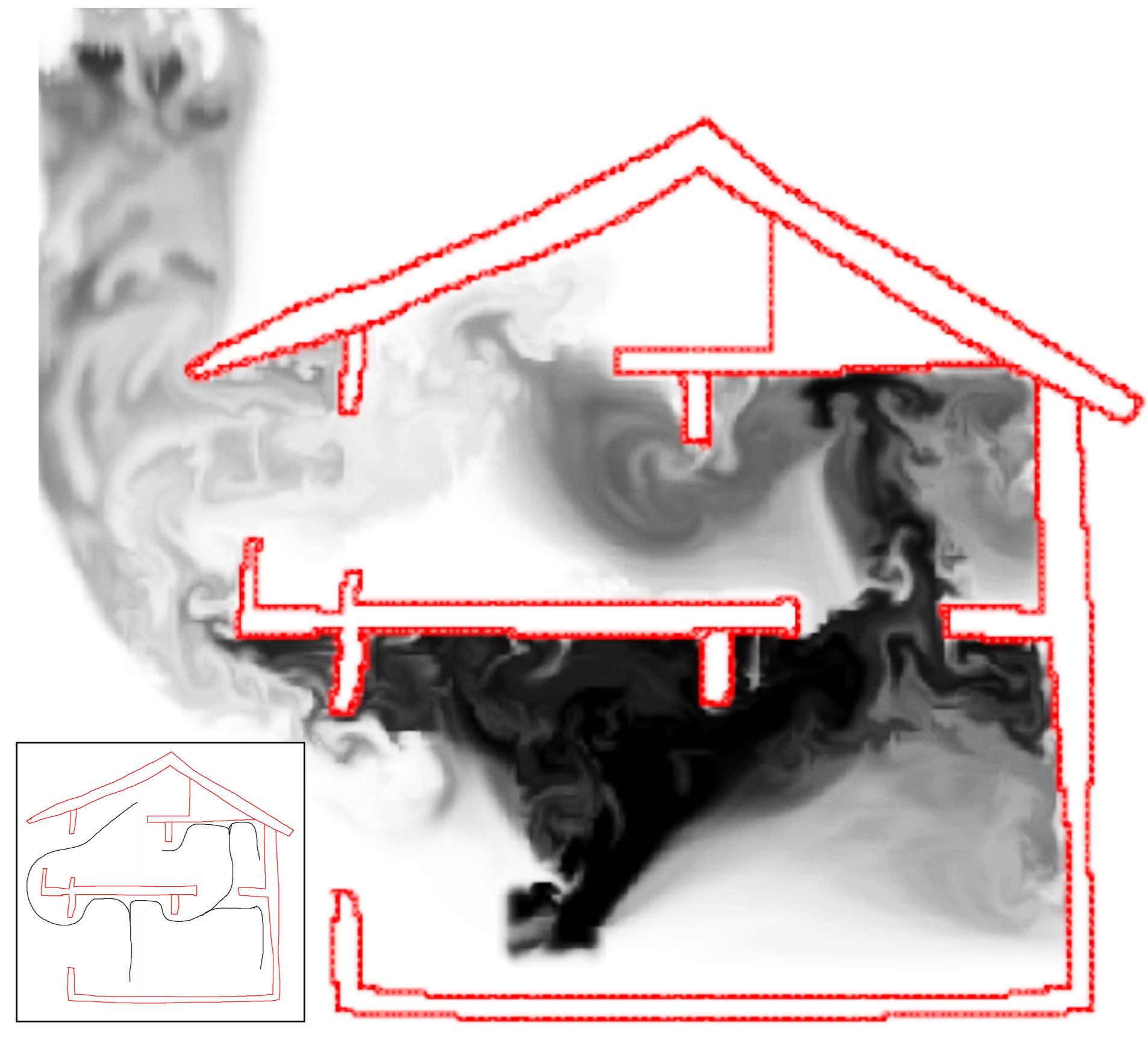}
\caption{Fire simulation of smoke movement inside a building. The bottom-left figure illustrates the user sketches with red lines for the building and black lines for smoke control.}
\label{fig:fire}
\end{figure}

The proposed approach can be utilized for education and demonstration purposes. For example, it is crucial to ensure the user safety in fire accidents with dense smoke. Figure~\ref{fig:fire} illustrates the smoke movement in fire to help the user seek the escape route in the building. With the help of the proposed design interface, the user can draw freely to control the smoke situation in real-time feedback. We believe the proposed system would be useful for dynamic illustrations and smoke simulations in various applications.

\section{System Evaluation}
The main objective of the proposed system is to help users in designing complex smoke simulations using the design interface. To verify the effectiveness of the proposed system, we evaluate the developed user interface using the designed user study and system evaluation. 

\subsection{User Study}
A user study was conducted to evaluate the system usability and user experience of the proposed system. The experiment was divided into the following \textit{Task A} and \textit{Task B} to measure the system usability and the design diversity.

\begin{figure}[b]
    \centering
    \includegraphics[width=0.95\linewidth]{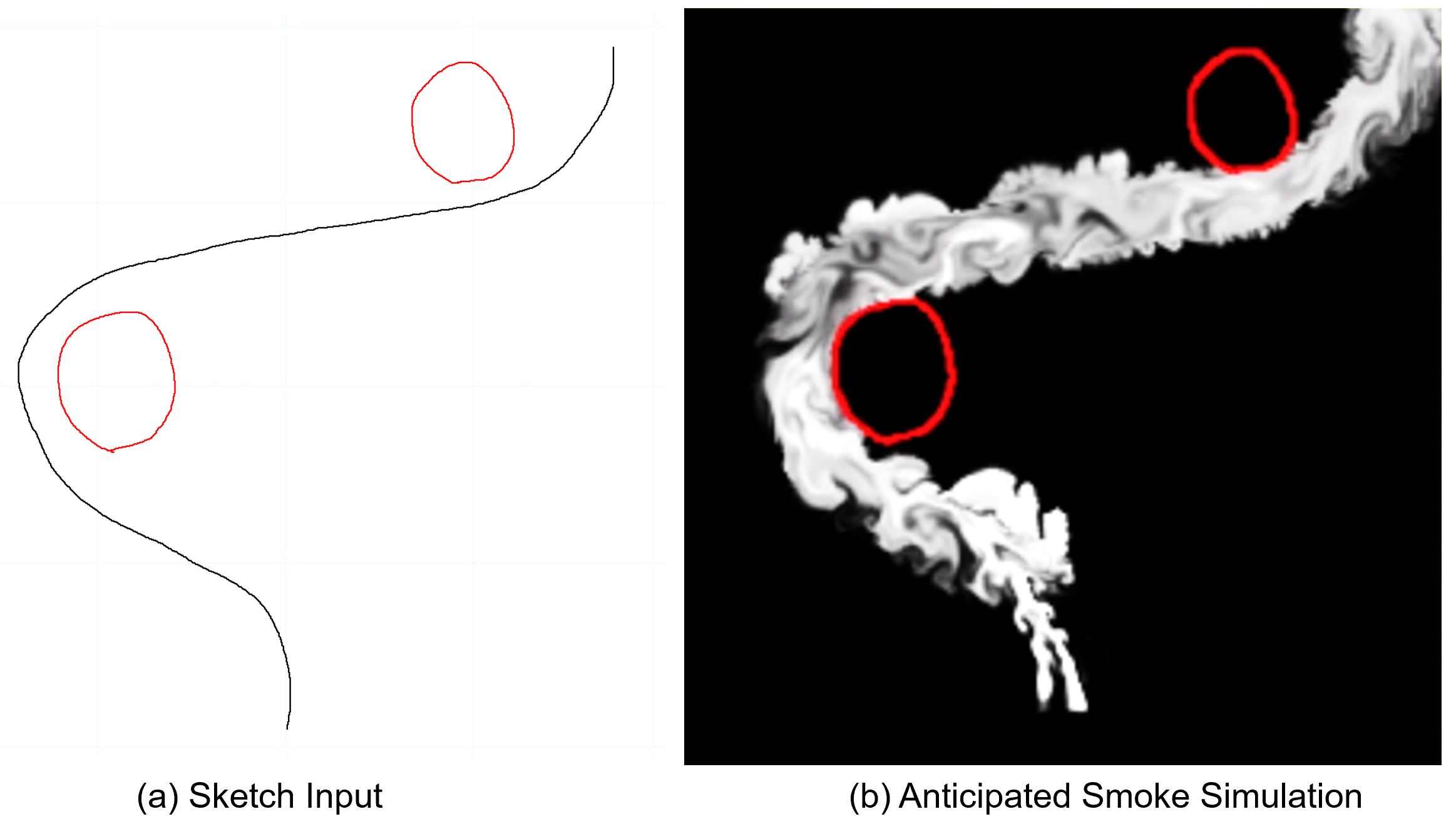}
\caption{The assigned design task A in our user study.}
\label{fig:expA-demo}
\end{figure}

For \textit{Task A}, the participants were asked to watch a smoke simulation that was created in advance. This smoke simulation could be viewed at any time during the experiment. The participants were asked to create a similar smoke simulation. Figure~\ref{fig:expA-demo} shows the sketch image and one frame image of the smoke simulation used in \textit{Task A}. For \textit{Task B}, the participants were asked to create a smoke simulation freely. 

We asked 20 graduate students to join the user study \cvmred{with the diverse majors in computer graphics, industry design, human-computer interaction, biology and chemistry} (10 males and 10 females, aged 20–35). The experimental procedure was organized as follows. We introduced the NASA-TLX tasks in 2 minutes because the participants were not familiar with the evaluation. We then explained the proposed system and allowed system practice for 5 minutes. After that, the participants completed \textit{Task A} and \textit{Task B} in turn with a time limit of 3–10 minutes for each task (20 minutes in total). The first 3 minutes were set as trial-and-error time to prevent the participants from completing the task without engaging. After the task experiment, we conducted a questionnaire survey using the system usability tool called the System Usability Scale (SUS) and the subjective assessment tool called NASA-TLX~\cite{brooke1996sus,hart1988development}. In addition, we collected user feedback as free comments in 5 minutes.

\begin{figure*}[t]
    \centering
    \includegraphics[width=0.9\linewidth]{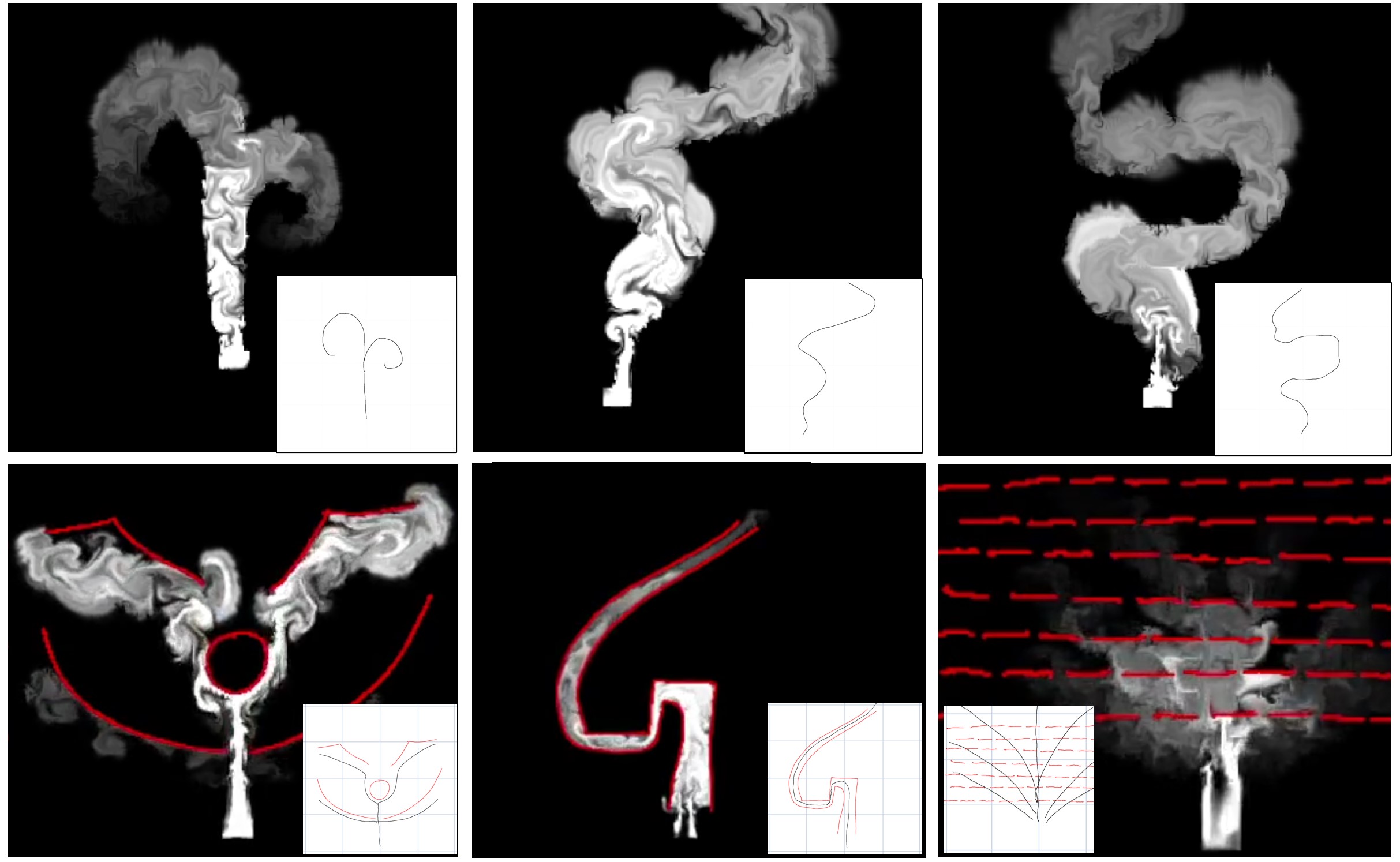}
\caption{The designed smoke simulations from user sketches (bottom right in each figure) \cvmred{in both simple and complex cases from} our user study.}
\label{fig:result-sketch}
\end{figure*}

\subsection{Evaluation Results}

In our user study, we verified the proposed design interface with both system usability and workload assessments. Figure~\ref{fig:result-sketch} shows three user sketches and the corresponding simulation results (from left to right, scaling parameter $c=1.4, 1.0, 1.8$) by the participants in our user study. Basically, a sketch drawn from the bottom to the top can generate the smoke flows without the disturbed velocity field.

\begin{figure}[t]
    \centering
    \includegraphics[width=0.9\linewidth]{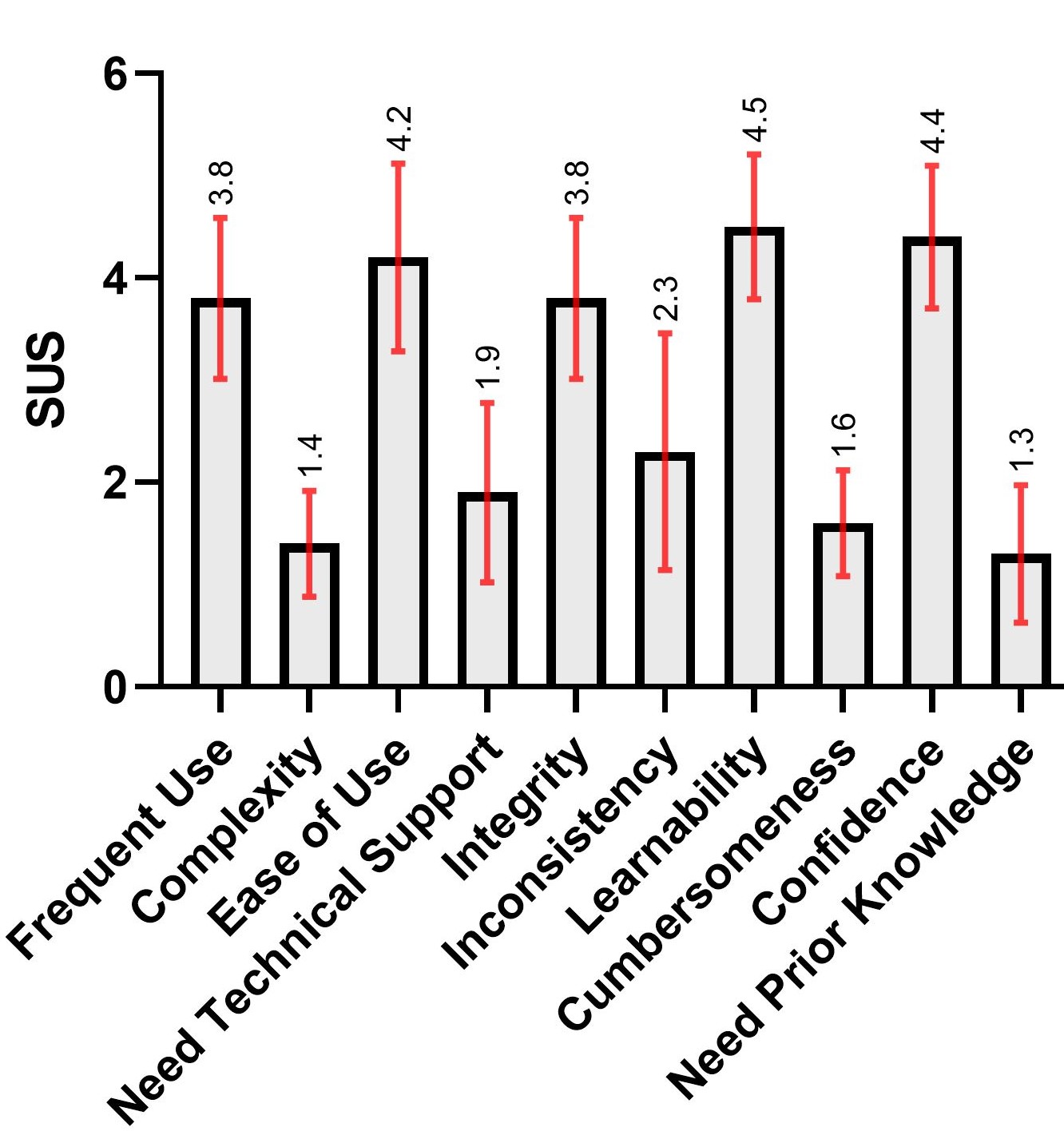}
    \caption{Results of SUS responses.}
    \label{fig:SUS}
\end{figure}

\subsubsection{System Usability}
The results of the SUS evaluation are shown in Figure~\ref{fig:SUS}, where the mean score is 80.5 (60 for min value, 95 for max). Because the average SUS score for a user interface was 76.2~\cite{bangor2009determining}. Therefore, this indicates that the proposed interface has excellent system usability.

For the SUS questionnaire, each question was answered using 5-point Likert scale (1 for strongly disagree, 5 for strongly agree). We had low scores for complexity, need technical support, cumbersomeness, and need prior knowledge. This indicated that the proposed interface is easy to use without the requirement of any expert knowledge. We had high scores for frequent use, ease of use, integrity, learnability, and confidence. This indicated that the users felt confident and comfortable using the proposed system. Note that we had a lower score for inconsistency which may be caused by the separated design and simulation windows. We plan to implement the integrated simulation window as part of future work. In terms of the user feedback, one participant reported that ``I did not understand why the smoke control is necessary after drawing obstacle". This suggests that the naturalness of the smoke design may be related to both the environment design and smoke control.

\begin{figure}[t]
    \centering
    \includegraphics[width=0.9\linewidth]{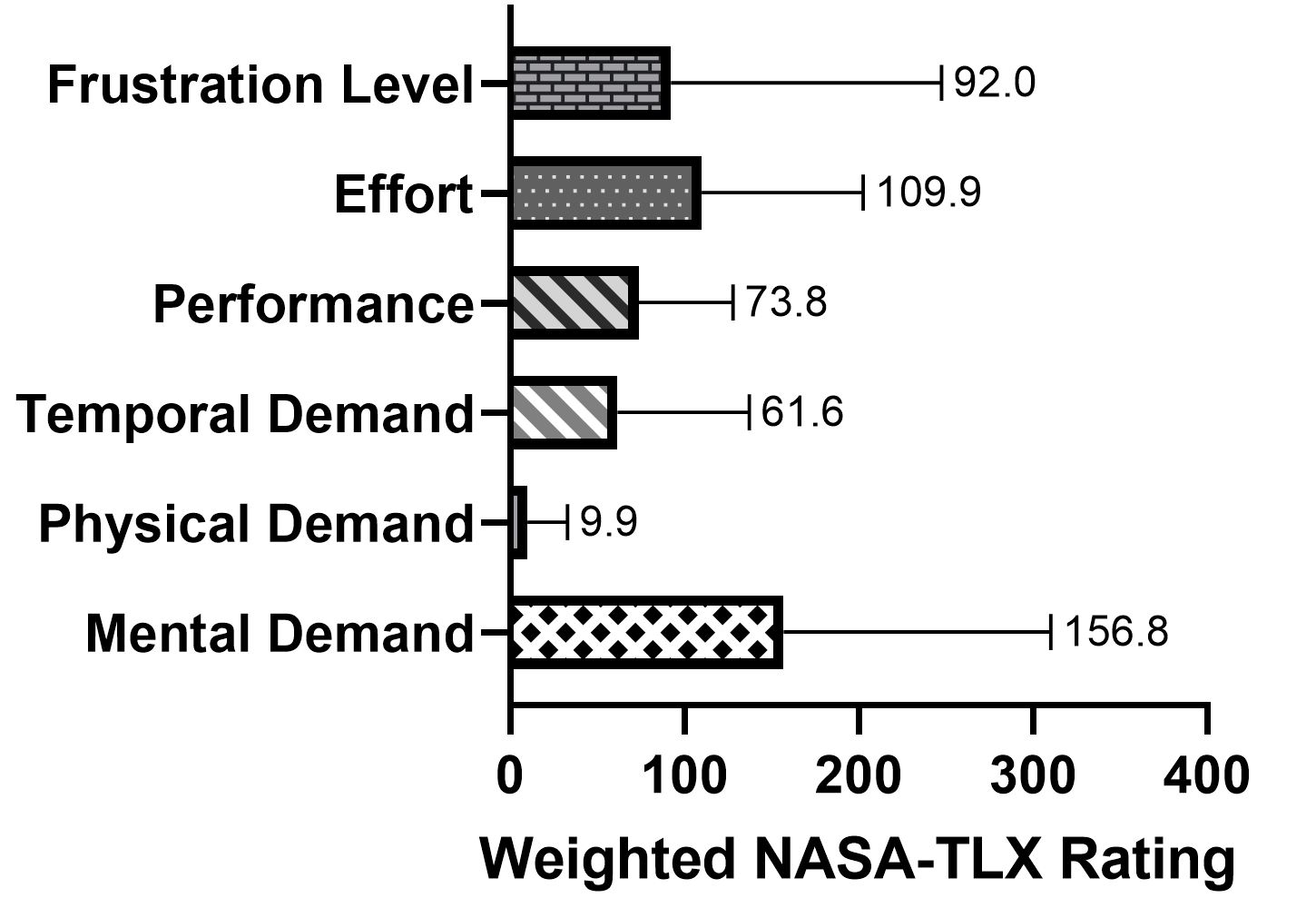}
    \caption{Results of weighted NASA-TLX rating evaluations.}
    \label{fig:NASA}
\end{figure}

\subsubsection{Subjective Assessment}
For the subjective NASA-TLX assessment, the highest rated item was mental demand as shown in Figure~\ref{fig:NASA}. Our user study involved the creation of a specified task and a free design task using the proposed system, and this may be the reason that the mental load of thinking was high. The physical demand and the temporal demand were rated as low, which indicates that the proposed sketch-based interface is easy to use. The average time costs for the two assigned tasks were around 6.1 minutes for \textit{Task A} and 6 minutes for \textit{Task B}.

In terms of the degree of performance and the effort required for the tasks, the weighted average score is not high due to the clarity of the proposed editing interface. Some participants commented that the degree of performance was related to the mental load. We thought that the interactive design interface allowed the participants to easily judge the quality of the drawing performance. Meanwhile, the low frustration level indicated the proposed interface is user friendly.  

\section{Conclusions}
In this work, we proposed \cvmred{DualSmoke}, a sketch-based design method for smoke \cvmred{illustration design} with a two-stage generative model. By extracting synthetic sketch data from flow patterns, the proposed system can generate the guiding force from the hand-drawn sketches for pattern-guided smoke simulation. The proposed method can allow common users without expert knowledge to edit fluid simulations easily. \cvmred{In ablation study,} we compared the proposed approach with the one-stage generative model to verify the proposed network structure, and the proposed system can obtain better global flow patterns and local turbulent details \cvmred{in the generated velocity fields}. Our user study and SUS \cvmred{evaluation} verified the proposed editing interface has good system usability, and the NASA-TLX
evaluation verified the workload is low.

In the current system, the user sketches were limited to an upward flow direction. It is possible to design a flexible design system with a training dataset for flows in multiple directions. In terms of geometry learning, normal maps from 2D sketches can be generated with generative models~\cite{su2018interactive,he21}. A similar approach can be used to estimate the geometry features of smoke simulations. This work calculated the guiding force from the generated results, and it is feasible to apply the guiding force fields to high-resolution simulation results~\cite{yuan2011pattern}.

\cvmred{Although this work aimed 2D smoke illustration design for common users without professional skills and domain knowledge}, we consider that an extension to 3D simulations is \cvmred{a promising research direction for visual effects}. With the training dataset for 3D smoke simulations, 3D FTLE~\cite{garth2007efficient} is a feasible solution, and 3D sketching would be useful for authoring this task, such as in an immersive virtual reality environment.

% \vspace{-15pt}
\subsection*{Acknowledgements}
We would like to thank the anonymous reviewers for their valuable comments. We thank the all participants in the user study. This work was supported by JAIST Research Grant, JSPS KAKENHI grant JP20K19845, Japan.

%%%%%%%%% REFERENCES
{\small
\bibliographystyle{ieee_fullname}
\bibliography{egbib}
}

\end{document}